\documentclass{elsart}

\usepackage{graphicx}
\usepackage{citesort}
\usepackage{color}

\begin{document}

\def\ov#1{\overline{#1}}

\begin{frontmatter}

\title{Recurrence intervals between earthquakes strongly 
depend on history}

 
\author[biu]{V.~Livina},
\author[biu]{S.~Tuzov},
\author[biu]{S.~Havlin},
\author[bu]{A.~Bunde}

\address[biu]{Minerva Center and Department of Physics, 
Bar-Ilan University, \break Ramat-Gan 52900, Israel}

\address[bu]{Institute f\"ur Theoretische Physik III,
Justus-Liebig-Universit\"at Giessen, 
Heinrish-Buff-Ring 16, 35392 Giessen, Germany}

\begin{abstract}
We study the statistics of the recurrence times between earthquakes
above a certain magnitude $M$ in California. We find that the 
distribution of the recurrence times 
strongly depends on the previous recurrence time $\tau_0$. As a consequence, 
the conditional mean recurrence time
$\hat \tau(\tau_0)$ between two 
events increases monotonically with $\tau_0$. For 
$\tau_0$ well below the average recurrence time 
$\ov{\tau}, \hat\tau(\tau_0)$ is 
smaller than $\ov{\tau}$, while for 
$\tau_0>\ov{\tau}$, $\hat\tau(\tau_0)$ is greater than
$\ov{\tau}$. Also the mean residual time until the next earthquake 
does not depend only on the elapsed time, but
also strongly on $\tau_0$. The larger $\tau_0$ is, the larger is the mean 
residual time. The above features should be taken 
into account in any earthquake prognosis.
\end{abstract}

\end{frontmatter}

Recently, Corral~\cite{Corral_PRL,Corral_PRE} 
studied the recurrence of earthquakes above a 
certain magnitude threshold $M$
in spatial areas delimited by a window of $L$ degrees in longitude 
and $L$ degrees in latitude. He found that the distribution
$D(\tau)$ of recurrence times $\tau$ scales with the mean recurrence 
time $\ov{\tau}$ as 
$$
D(\tau) = {{1}\over {\ov{\tau}}}\,f\,(\tau/\ov{\tau}),
$$
where the function $f(\Theta)$ is quite universal and independent of 
$M$. For $\Theta$ below 1, 
$f$ can be approximated by a power-law, while
for $\Theta\gg 1,\ f(\Theta)$ decays exponentially with 
$\Theta$. 
As a consequence of the deviation from a Poissonian 
decay~\cite{Sornette}, 
the mean residual time to the next event increases with the elapsed 
time~\cite{Corral_PRL,Corral_PRE}. 

In this paper, we study the statistics of the recurrence intervals of the 
California database~\cite{web}
and find that both quantities, the recurrence interval
distribution $D(\tau)$ and the mean residual time to the next
earthquake strongly depend 
on the previous recurrence time interval $\tau_0$.

We study the records from the local Californian earthquake 
catalog~\cite{web} for the 
period 1981-2003 in the area 30.5-38.5N latitude, 114-122W longitude,
with minimal magnitude 
threshold value $M=2$ and minimal recurrence times 2 mins~\cite{note}.
Similarly to Corral [1], we consider the earthquakes  in the region that are above a certain threshold $M$, as a linear process 
in time $\{t_i\}$ without taking into account 
the spatial coordinates of the event hypocenters. 
We are 
interested in the recurrence intervals  $\tau_i=t_i-t_{i-1}$  between these earthquakes.

In records without memory, the (conditional) distribution 
$D(\tau\vert\tau_0)$  of  recurrence intervals $\tau$ 
that directly follow 
a certain interval $\tau_0$, does not depend on the value of $\tau_0$
and is identical to $D(\tau)$.
In contrast, in records with long-term memory, there is a pronounced dependence
of $D(\tau\vert\tau_0)$  on  $\tau_0$ [7,8]. 
To study possible memory effects in the earthquake records with a reliable statistics,
 we have studied the
conditional distribution $D(\tau|\tau_0)$ not for a specific $\tau_0$ value, but for
values of $\tau_0$ in certain intervals.
To this end,
we have sorted the record of $N$ recurrence intervals 
in increasing order and divided it
into four subrecords $Q_1$, $Q_2$, $Q_3$ and $Q_4$, such that 
each subrecord contains one quarter of the total number of recurrence intervals.
By definition, the $N/4$ lowest recurrence intervals are in $Q_1$, while the $N/4$ 
largest intervals are in $Q_4$.

\begin{figure}[h!]
\centerline{\includegraphics[width=0.7\textwidth]{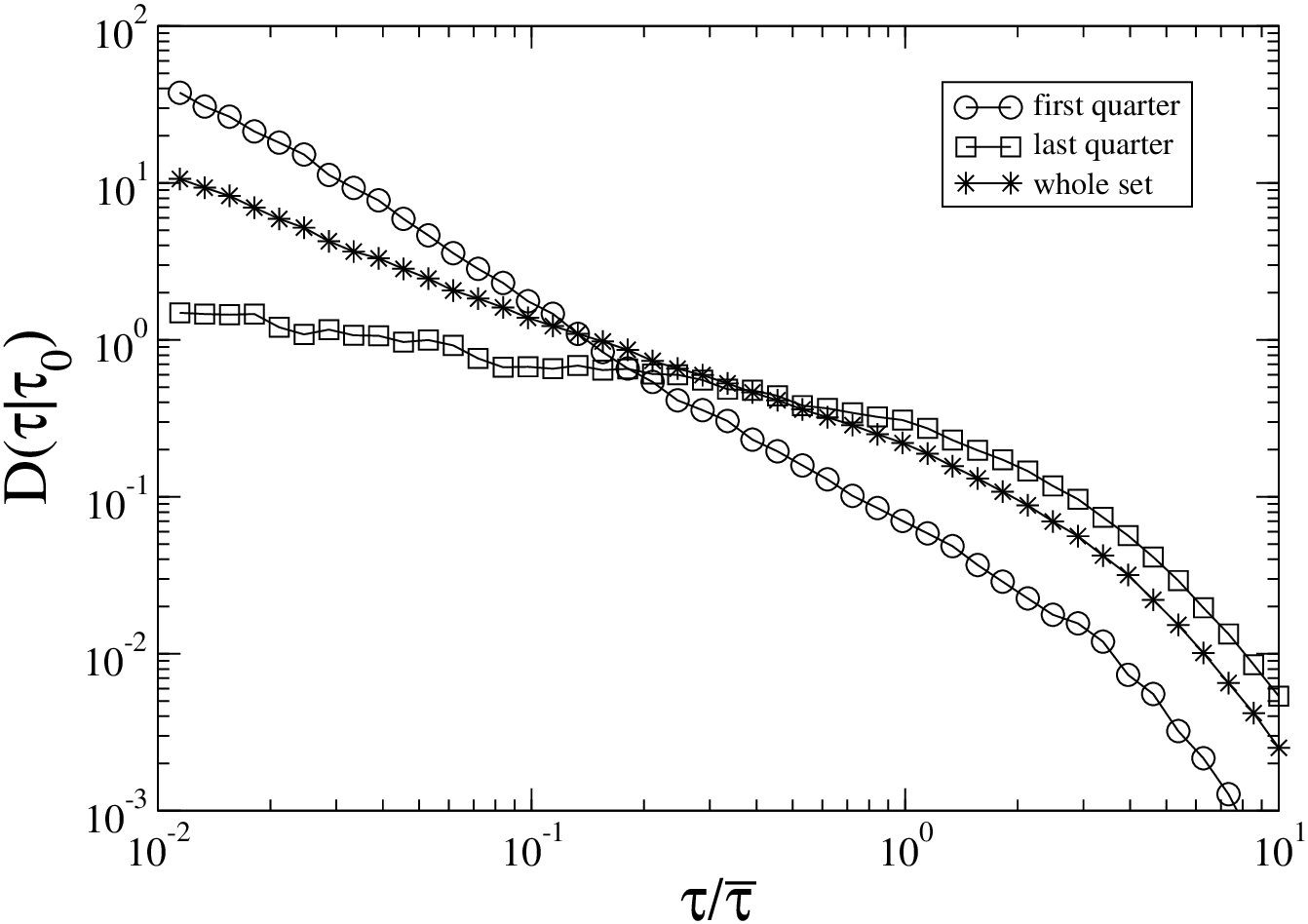}}
\caption{Conditional probability distribution for 
the recurrence time intervals between earthquakes above a 
threshold $M\simeq 2.5$ following 
recurrence time $\tau_0$ from the first quarter (circles) and the last quarter 
(squares) of the recurrence time, and the unconditional probability 
(stars). 
To improve statistics,
averages 
were taken for $2.25 \le M\le 2.75$.
}
\label{fig1}
\end{figure}

Figure~\ref{fig1} shows $D(\tau|\tau_0)$ for 
$\tau_0$ averaged over $Q_1$ and $Q_4$. For comparison, we also show the
unconditional distribution function $D(\tau)$. 
To improve the statistics, we used logarithmic binning. We considered time scales from 2~minutes to $10\ov{\tau}$,
with 50 log-bins,  counted
the number of recurrence intervals within each bin
and divided it by the size of the bin.  
To further improve the statistics, 
we averaged the probability distribution 
over threshold values $M=2.25\,\dots\,2.75$   
around $M\simeq 2.5$.
Finally, we normalized the probability distribution to obtain the probability
densities of interest.
The figure shows that for $\tau$ well below its mean value $\ov{\tau}$, 
the probability of finding $\tau$ below (above) $\ov{\tau}$ is enhanced 
(decreased) 
compared with $D(\tau)$
for $\tau_0$ in $Q_1$,
while the opposite occurs 
for $\tau_0$ in $Q_4$.

\begin{figure}[h!]
\centerline{\includegraphics[width=0.7\textwidth]{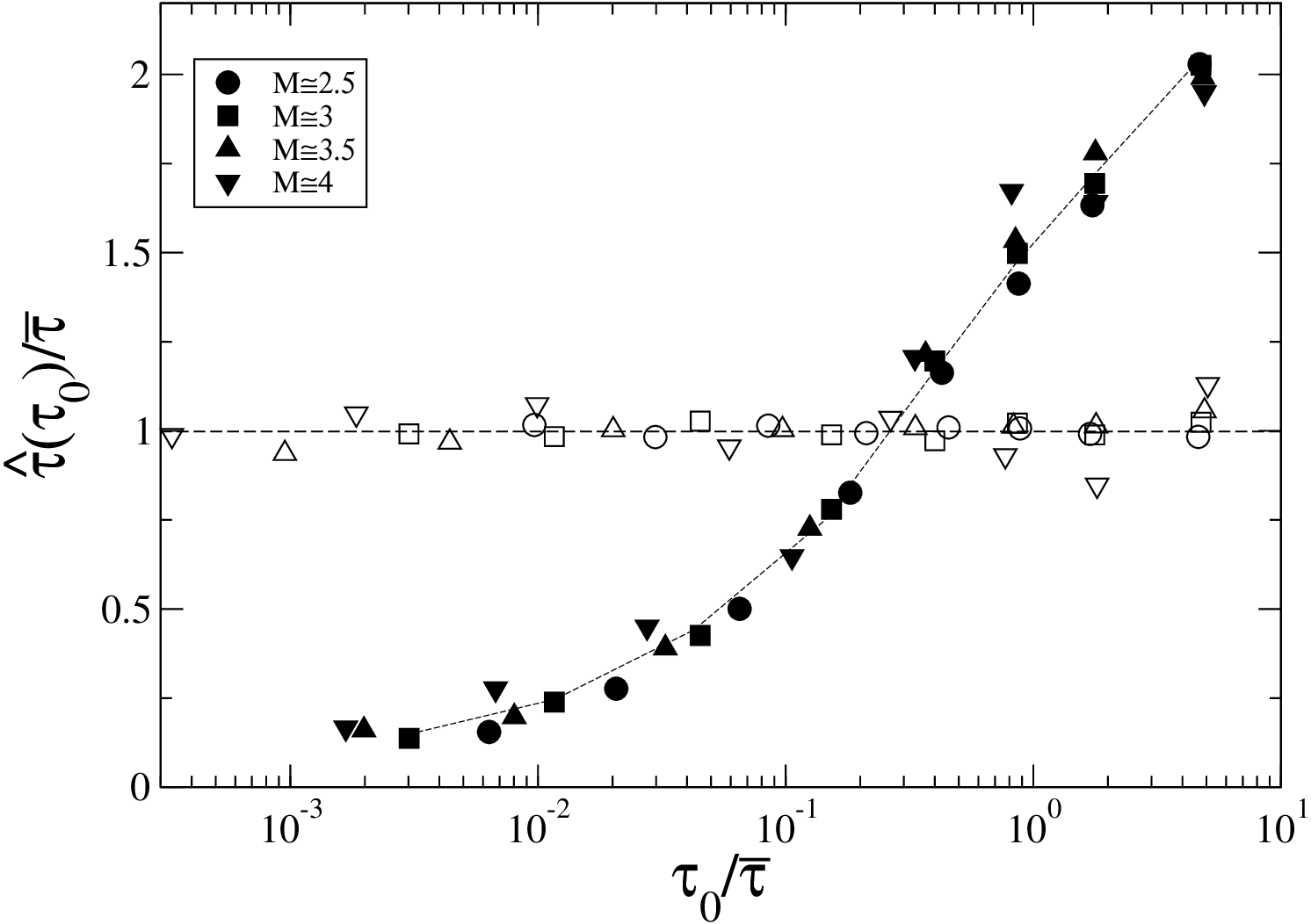}}
\caption{Expected recurrence time 
$\hat\tau(\tau_0)$
between earthquakes
above thresholds $M\simeq 2.5$ (full circles), $M\simeq 3$ (full squares),
$M\simeq 3.5$ (full triangles up), and $M\simeq 4$ (full triangle down)
following $\tau_0$ taken from the eight octaves 
described in the text. Averages 
are taken in intervals $M\pm 0.25$ to obtain better statistics. 
The open symbols represent the analysis of the randomly shuffled 
recurrence time record, yielding $\hat\tau(\tau_0)/\ov{\tau}\simeq 1$.
}
\label{fig2}
\end{figure}

By definition, $\hat \tau(\tau_0)$ is 
the mean recurrence intervals, when the two events before were separated by
an interval  $\tau_0$. 
The memory effect in the conditional distribution function 
$D(\tau|\tau_0)$ leads 
to an explicit dependence of 
$\hat\tau (\tau_0)$ on $\tau_0$.
To calculate $\hat \tau (\tau_0)$,
we divided the sorted (in increasing order) 
record of recurrence intervals into 8 consecutive octaves. Each octave
contains $N/8$ intervals. In each interval, we calculate the mean value. 
We studied 
$\hat \tau$ as a function of $\tau_0/\ov{\tau}$, where now
$\tau_0$ denotes  the mean recurrence time in the octave.
Figure~\ref{fig2} 
shows $\hat\tau(\tau_0)/\ov{\tau}$ as a function of $\tau_0/\ov{\tau}$ and
clearly demonstrates the strong effect of the memory. Small and large 
recurrence intervals are more
likely to be followed by small and large ones, respectively,  
$\hat \tau/\ov{\tau}$ is well below (above) one for 
$\tau_0/\ov{\tau}$ well below (above)
one. When the recurrence intervals are randomly shuffled (no memory),
we obtain $\hat\tau(\tau_0)/\ov{\tau} \cong 1$, see Fig.~\ref{fig2}, open 
symbols.

A more general 
quantity is the expected residual time 
$\hat\tau(x|\tau_0)$ 
to the next event, 
when time $x$ has been already elapsed. For 
$x=0$, $\hat \tau(0|\tau_0)$ 
is identical 
to $\hat\tau(\tau_0)$. In general, $\hat \tau(x|\tau_0)$ 
is related to $D(\tau|\tau_0)$ by
\begin{equation} 
\hat\tau(x| \tau_0)=\int_x^\infty (\tau -x) 
D(\tau| \tau_0)\,d\tau \left/
\int_x^\infty D(\tau| \tau_0)\,d\tau\right. . 
\end{equation}
For uncorrelated records, 
$D(\tau|\tau_0)$ is Poissonian, and 
$\hat\tau(x|\tau_0)/\ov{\tau}=1$.

\begin{figure}[h!]
\centerline{\includegraphics[width=0.7\textwidth]{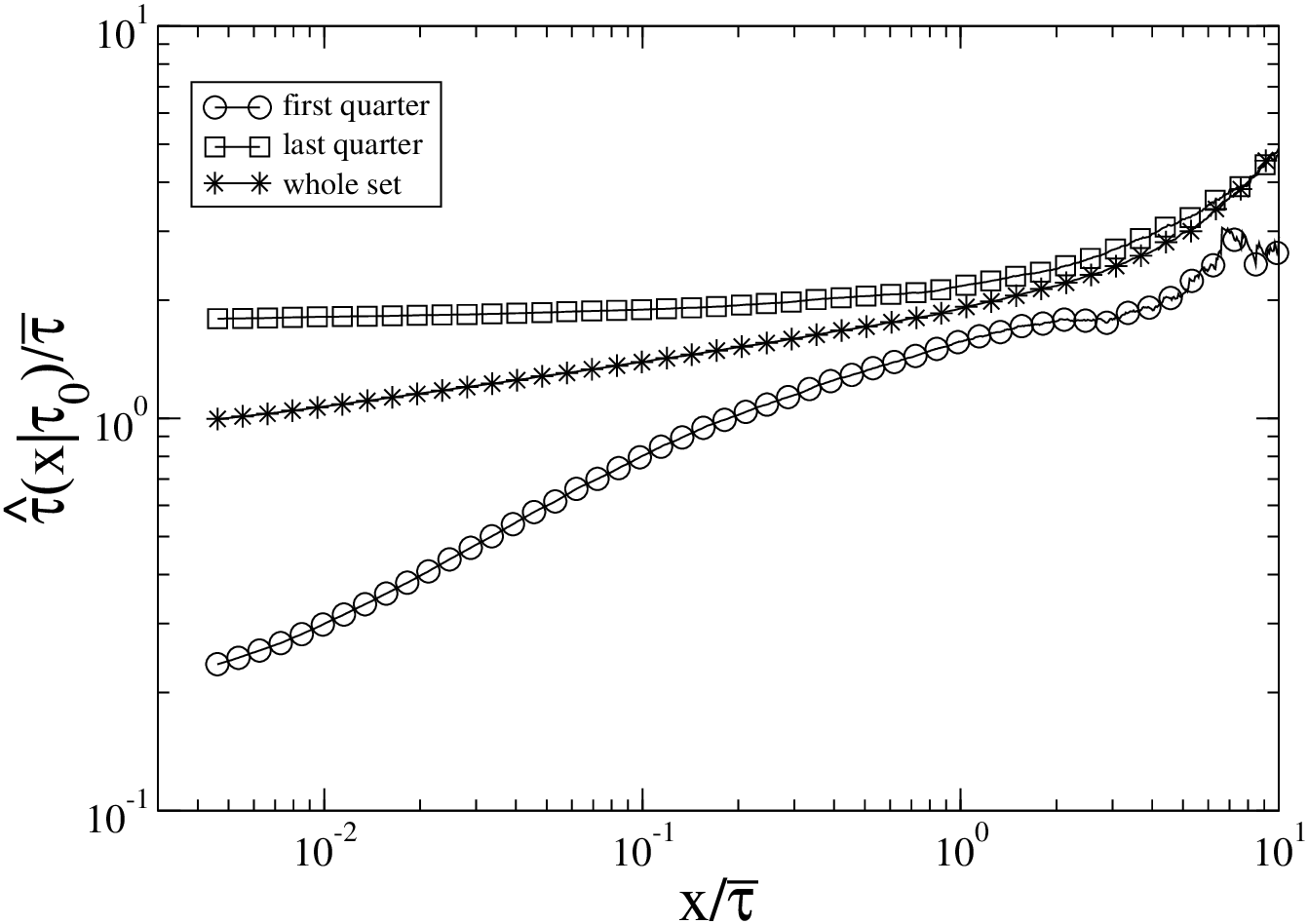}}
\caption{Conditional mean residual time 
to the next earthquake above a threshold $M\simeq 2.5$
following recurrence time $\tau_0$ taken from the first quarter 
(bottom curve) and the 
last quarter (top curve) of the recurrence intervals, and unconditional 
mean residual time (middle curve).
To improve statistics, average is taken for $2.25\le M \le 2.75$.
}
\label{fig3}
\end{figure}

Figure~\ref{fig3} clearly shows that 
$\hat\tau(x|\tau_0)$ depends on 
both $x$ and $\tau_0$.
With increasing 
$x$, the expected residual time to the next event increases, as is shown
in Fig.~\ref{fig3}, for values of $\tau_0$ from $Q_1$ and $Q_4$ 
(top and bottom curves). Thus, when $\tau_0$ increases, 
$\hat\tau(x|\tau_0)$ increases for all values of $x$.
The middle curve
shows the expected residual time averaged over all 
$\tau_0$, i.e. the unconditional 
residual time $\hat\tau(x)$.  In this case, the interval between the last two events
is not taken into account, and the slower-than-Poisson-decrease
of the unconditional distribution function $D(\tau)$, 
Eq.~(1), leads to the
anomalous increase of the mean residual 
time with the elapsed time~\cite{Sornette}.

Our results for the unconditional residual time function for 
Californian earthquakes are very similar to the results of 
Corral~\cite{Corral_prep} that were obtained for worldwide earthquake records.
As shown here, there is a strong memory in the earthquake recurrence 
intervals, which influences significantly the residual time.
Similar memory effects have been obtained recently for river flux, 
temperature and precipitation records  
(see~\cite{Bunde_PhysA,Bunde_prep}).

To summarize, we have studied the memory effect in the earthquake events and 
showed that the distribution of the recurrence times 
and the mean residual time until the next earthquake 
strongly depend on the previous recurrence time. 
The conditional mean recurrence time between two events monotonically increases 
with $\tau_0$. 
These results should be taken into account in an efficient risk
evaluation and forecasting of earthquakes.
It is very plausible that the origin of these effects is due to 
long-term
persistence in the earthquake occurence.

\vspace{10mm}

\begin{thebibliography}{}
\bibitem{note}
This condition is taken due to 
probable incompleteness of the catalog
in the very short-time scales (see~\cite{Corral_PRL}).
\end{thebibliography}
\end{document}